\documentclass[journal=cmatex,manuscript=article,layout=twocolumn]{achemso}
\setkeys{acs}{maxauthors = 0}
\usepackage[version=3]{mhchem}
\usepackage[T1]{fontenc}
\usepackage{color,ulem}
\usepackage{amsmath,amssymb}
\usepackage{textcomp}

\usepackage{color,soul} 
\usepackage[english]{babel}

\makeatletter
\let\l@addto@macro\relax
\makeatother
\usepackage[fontsize=10pt]{scrextend}

\usepackage{color,ulem}

\usepackage{ulem}
\graphicspath{{figures/}}

\let\oldmaketitle\maketitle
\let\maketitle\relax

\author{Anuj Goyal}
\affiliation{Colorado School of Mines, Golden, CO 80401, USA}
\author{Prashun Gorai}
\affiliation{Colorado School of Mines, Golden, CO 80401, USA}
\author{Shashwat Anand}
\affiliation{Northwestern University, Evanston, IL, USA}
\author{Eric S. Toberer}
\affiliation{Colorado School of Mines, Golden, CO 80401, USA}
\author{G. Jeffrey Snyder}
\affiliation{Northwestern University, Evanston, IL, USA}
\author{Vladan Stevanovi\'{c}}
\email{vstevano@mines.edu}
\affiliation{Colorado School of Mines, Golden, CO 80401, USA}

\title{On the dopability of semiconductors and governing materials properties}

\begin{document}

\begin{tocentry}
\includegraphics{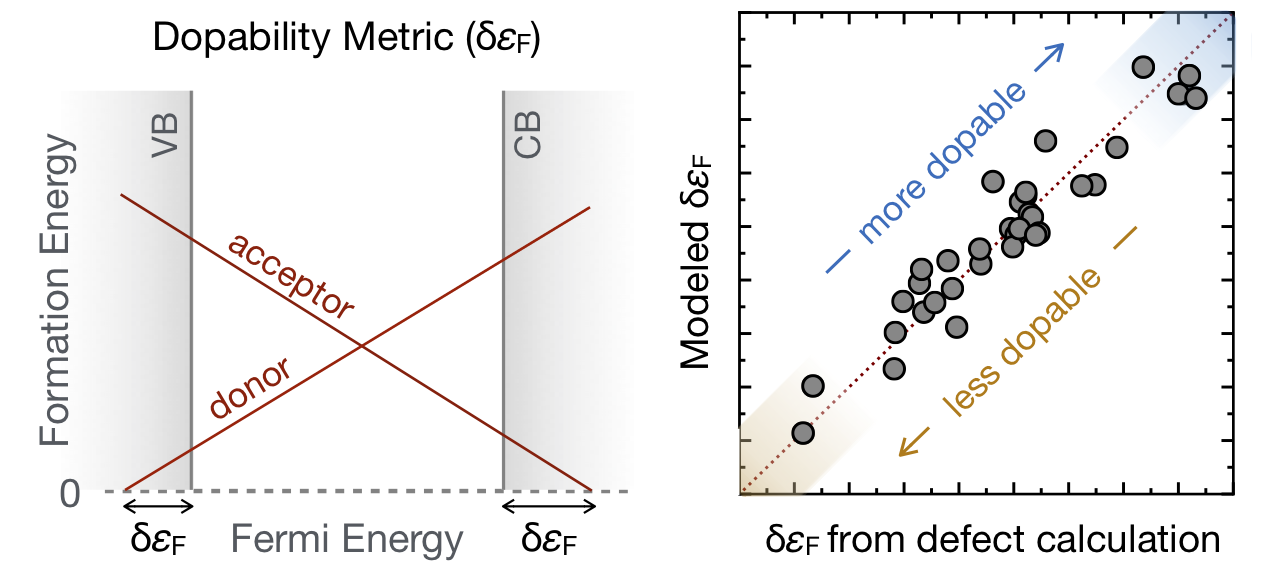}
\end{tocentry}

\twocolumn[
\begin{@twocolumnfalse}
\oldmaketitle
\begin{abstract}
To be practical, semiconductors need to be doped. Sometimes, to nearly degenerate levels, e.g. in applications such as thermoelectric, transparent electronics or power electronics. However, many materials with finite band gaps are not dopable at all, while many others exhibit strong preference toward allowing either $p$- or $n$-type doping, but not both. In this work, we develop a model description of semiconductor dopability and formulate design principles in terms of governing materials properties. Our approach, which builds upon the semiconductor defect theory applied to a suitably devised (tight-binding) model system, reveals analytic relationships between intrinsic materials properties and the semiconductor dopability, and elucidates the role and the insufficiency of previously suggested descriptors such as the absolute band edge positions. We validate our model against a number of classic binary semiconductors and discuss its extension to more complex chemistries and the utility in large-scale material searches.\\\\
\end{abstract}
\end{@twocolumnfalse}
]

\section{Introduction}\label{sec:intro}
\begin{figure}[t!]
\includegraphics[width=0.85\linewidth]{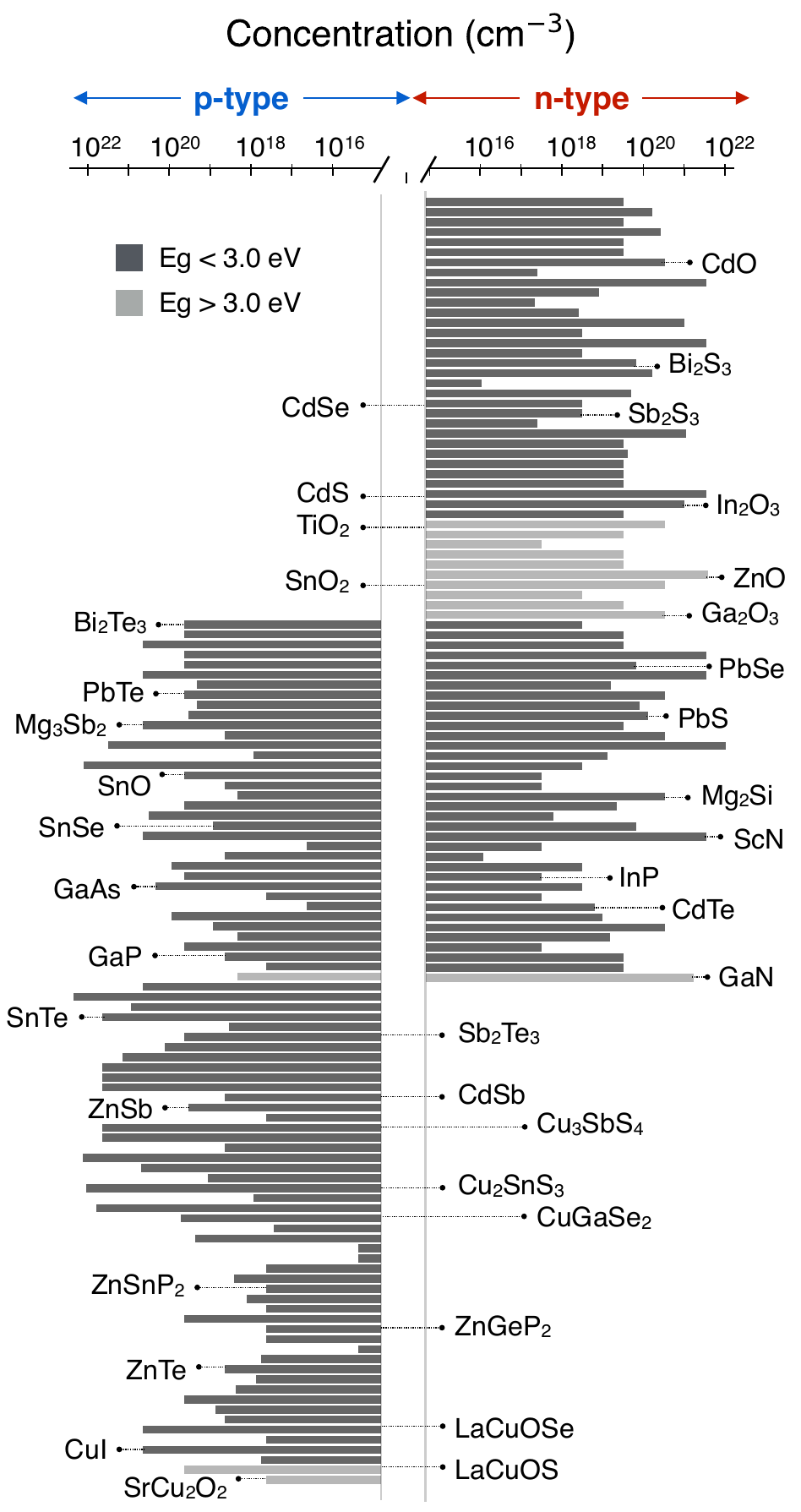}
\caption{\label{fig:1}Histogram of maximal reported charge carrier concentrations for various binary and ternary semiconductors. Light and dark shades of grey distinguish the band gap value. Data obtained from Refs.\cite{Burbano2011, Sachet2015, Dou2002, sam_npj:2018, Martinez2017, Biswas2012, Wu2003, Kumar2013, Ibuki2000, Kumaravel2011, Segev2005, Colston2018, Crandall1968, Walsh2008a, Limpijumnong2009a, vladan_PCCP:2014, Kim2012, VanBenthem2001, Tufte1967, Dimitrievska2016, Ueda1992, Kawazoe2010, Ozgur2005a, Nagasawa1966, Edwards2004, Bellal2009, Hadjarab2007, Pearton2018a, Lany2018, Yashima1998, Filip2013, Hor2009, Goyal2017b, Ohno2018, Shuai2015, Zhang2017, Varley2013, Hosono2011, Strauss1961, Kolezynski2015, Nolas2007, Duong2016, Kumagai2018, Deng2015, Saha2017, Fioretti2016, Vidal2012b, Xiao2015, Bugajski1985, Neave1983, Sun1991, Su2008, Segall1963, Nagaoka2018, Simpson1992, VanDeWalle1993, Ueno2017, Tsu1968, Tenga2009, Hruby1985, Bottger2011, Martinez2017, Liu2014a, Chen2017, Thompson1989, Heinemann2013, Lindberg2018, Ueda2004, Barati2009, Hiramatsu2003, Ahn2016, Yang2016, Tate2009, Kawazoe1997, Hiramatsu2003, Ohta2002, Kudo1998} and for details see Table~S1 in the SI.}
\end{figure}

The ability of classic semiconductors such as Si, GaAs, and PbTe to be doped both $p$- and $n$-type and to nearly arbitrary charge carrier concentrations is an exception rather than a rule. 
This is clearly illustrated by our literature survey, depicted in Fig.~\ref{fig:1}. We collected the highest measured charge carrier concentrations for about 130 binary and ternary semiconductors. A number of these compounds have relatively low (maximal) reported charge carrier concentrations and only a small fraction (36 out of 130) have been successfully doped both $p$- and $n$-type. These numbers decrease significantly as the material band gap increases such that there are very few wide-gap semiconductors that are dopable at all (13 out of 130 with gaps above 3 eV), with GaN being the only (weakly) ambipolar semiconductor with the band gap exceeding 3 eV. Recent predictions offer some hope that ambipolar wide-gap materials could exist, though experimental validation is still needed \cite{scanlon_CM:2017, varley_CM:2017, varley_PRB:2014}. Majority (95 out of 130) of the compounds from Fig.~\ref{fig:1} likely suffer from the so-called doping asymmetry, meaning that they can be doped either $p$- or $n$-type, but not both. 
%

GaN is a notable example of how finding ways to overcome these doping tendencies (or bottlenecks) can be transformative. A nominally exclusively $n$-type semiconductor was successfully doped $p$-type via a rather unconventional non-equilibrium processing route that allows insertion of acceptor behaving Mg substitutional impurities in much higher concentrations than possible under equilibrium conditions. This accomplishment enabled the development of a blue light emitting diode which was awarded 2014 Nobel prize in physics \cite{akasaki_RMP:2015, amano_RMP:2015, nakamura_RMP:2015}. Likewise, numerous attempts have been made to dope ZnO, SnO$_2$, and In$_2$O$_3$ $p$-type \cite{Anderson2009, Hosono2011a, King_2011}; however, with very little success, and to this day, these compounds are regarded as exclusively $n$-type semiconductors \cite{lany_PRB:2009, scanlon_JMC:2012}. Recent predictions suggest that the $p$-type doping in ZnO could be attained; however, not in the ground state wurtzite structure but in the metastable, high-pressure rocksalt phase \cite{goyal_PRM:2018}.  Another important (counter) example is Mg$_3$Sb$_2$, which was for long regarded as an exclusively $p$-type semiconductor, a belief that was recently contested by the successful (equilibrium) $n$-doping followed by the demonstration of high thermoelectric performance in the $n$-type Mg$_3$Sb$_2$ \cite{tamaki_AM:2016, saneyuki_Joule:2018}. 
These doping tendencies and bottlenecks represent a critical obstacle for the discovery and design of novel functional materials; especially for applications such as thermoelectric, transparent, and power electronics where achieveing nearly degenerate charge carrier concentrations is of utmost importance \cite{Alberi_2018,gorai_NRM:2017,Gorai_EES:2019}. 

\begin{figure}[t!]
\includegraphics[width=\linewidth]{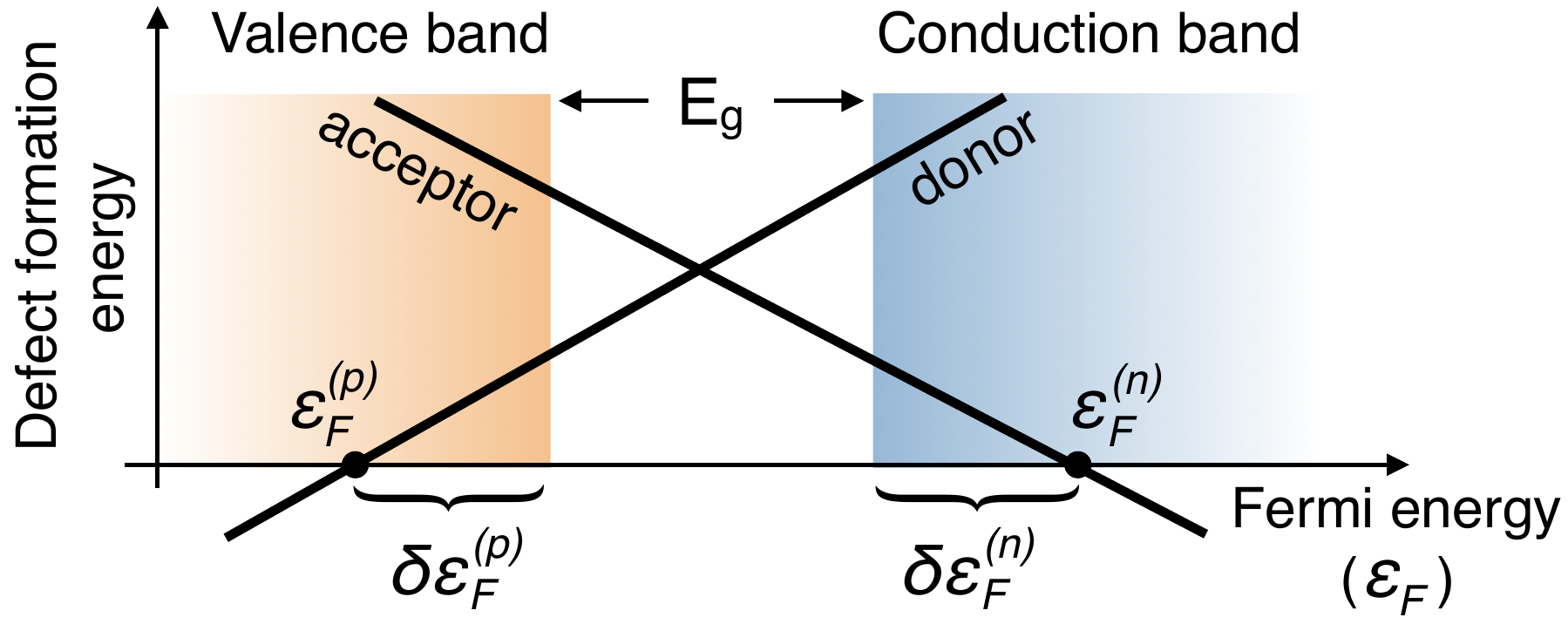}
\caption{\label{fig:2}
Schematic of the defect formation energy dependence on the Fermi energy for donor and acceptor defects. The ``doping pinning energies'' that determine the dopability of a material are denoted as $\varepsilon_F^{(n)}$ and $\varepsilon_F^{(p)}$. In our work we define dopability metrics $\delta \varepsilon_F^{(n)}$ and $\delta \varepsilon_F^{(p)}$ shown in the figure as deviations of the pinning energies from the corresponding band edge (see text for details). }
\end{figure}

Dopability of III-V and II-VI semiconductors  was investigated previously. Zunger formulated his practical doping principles as related to the formation of intrinsic compensating defects \cite{zunger_APL:2003}. Namely, the formation energy of any acceptor defect exhibits decreasing linear dependence on the Fermi energy ($\varepsilon_F$) while for the donors it increases linearly with $\varepsilon_F$ as shown in Fig.~\ref{fig:2}. Hence, there will be a special $\varepsilon_F^{(n)}$ value above which the energy to form intrinsic acceptors becomes negative (exothermic). Similarly, there is a special $\varepsilon_F^{(p)}$ below which the formation of intrinsic donors will be exothermic. As a consequence, any attempt to dope the system $n$-type by increasing $\varepsilon_F$ beyond the $\varepsilon_F^{(n)}$ will be met with the opposition in the form of spontaneous formation of intrinsic acceptor (electron-compensating) defects. If $\varepsilon_F^{(n)}$ occurs near or inside the conduction band, the system will allow introduction of electrons and hence, $\varepsilon_F^{(n)}$ represents a natural upper limit for $n$-type doping. Analogously, intrinsic donors and the resulting $\varepsilon_F^{(p)}$ determine the limit for $p$-type doping. 

What Zunger noticed is that within III-V and II-VI semiconductors the $n$- and $p$-type pinning energies $\varepsilon_F^{(n)}$ and $\varepsilon_F^{(p)}$, as he called them, obtained from defect calculations approximately align. This implies that in order to be dopable, the semiconductor band edges need to be close to these ``universal" pining energies. That is, lower the CBM position relative to the $n$-pinning energy, the more $n$-type dopable the systems is, and conversely, higher the VBM relative to the $p$-pinning energy, the more $p$-type dopable the semiconductor. While certainly practical, Zunger's doping principles critically rely on the empirically observed alignment of the doping pinning energies, which, as we will show later, does not hold generally for compounds outside III-V and II-VI semiconductors.   

Similar design principles emerge from the consideration of another ``universal'' reference, the branch point energy \cite{schleife_APL:2009}. It is defined as the energy at which the electronic states at the surface and/or interfaces change their character from predominantly VB-like to mostly CB-like, and is also used implicitly to define dopability as related to the proximity of the band edges. Lower the CBM relative to branch point energy, more $n$-type dopable the system is; and higher the VBM, more $p$-type dopable the system is. If the branch point energy occurs close to the mid-gap, the system could be either ambipolar dopable or insulating. The expected universal alignment of the branch point energies between different materials then implies dopability design principles in terms of the positions of the band edges similar to those proposed by Zunger, but now relative to this different reference. However, recent work shows that considerations based on the branch point energy can be used to identify $n$-type dopable systems much better than the ambipolar or the $p$-type systems \cite{rachel_CM:2018}. 

%

Also recently, Miller \textit{et al.} used machine learning to develop an empirical model for dopability in diamond-like semiconductors \cite{sam_npj:2018}. While the developed model shows remarkable accuracy in reproducing and predicting achievable carrier concentrations, as with any machine learning model its relation to the underlying physics is unclear (not causal) and the transferability beyond the diamond-like semiconductors is questionable due to the scarcity of measured carrier concentrations needed for the model development.

Herein, we revisit the problem of predicting dopability of semiconductors and build upon these previous works. We ask the question of the complete set of governing intrinsic material properties and the causal relationship between them, without making the largely qualitative and heuristic assumptions about the alignment of the pinning energies or the connection of dopability to the branch point energy. 
We do this by formulating a model description of a binary ionic semiconductor using the tight-binding model for non-interacting electrons supplemented by the nucleus-nucleus repulsion (pair) potential. We use this model to derive analytic expressions for the formation energy of intrinsic donors and acceptors and the associated doping pinning energies. The model is validated against directly calculated (first-principles) pinning energies. Finally, we analyze the new insights that are provided by the model, the role of previous heuristics as well as its utility in searching for dopable materials. It is important to note however, that modern defect theory and defect calculations \cite{freysoldt_RMP:2014, Lany_2009} can be used to predict both intrinsic limits to dopability (doping pinning energies) and the effectiveness of extrinsic dopants \cite{saneyuki_Joule:2018,gorai_JMCA:2018}. The aim here is to uncover physical principles of semiconductor dopability that are usually implicit and hidden in numerical approaches. 

\section{Dopability model}
%
\subsection{Construction}\label{ssec:model}
\begin{figure}[t!]
\includegraphics[width=\linewidth]{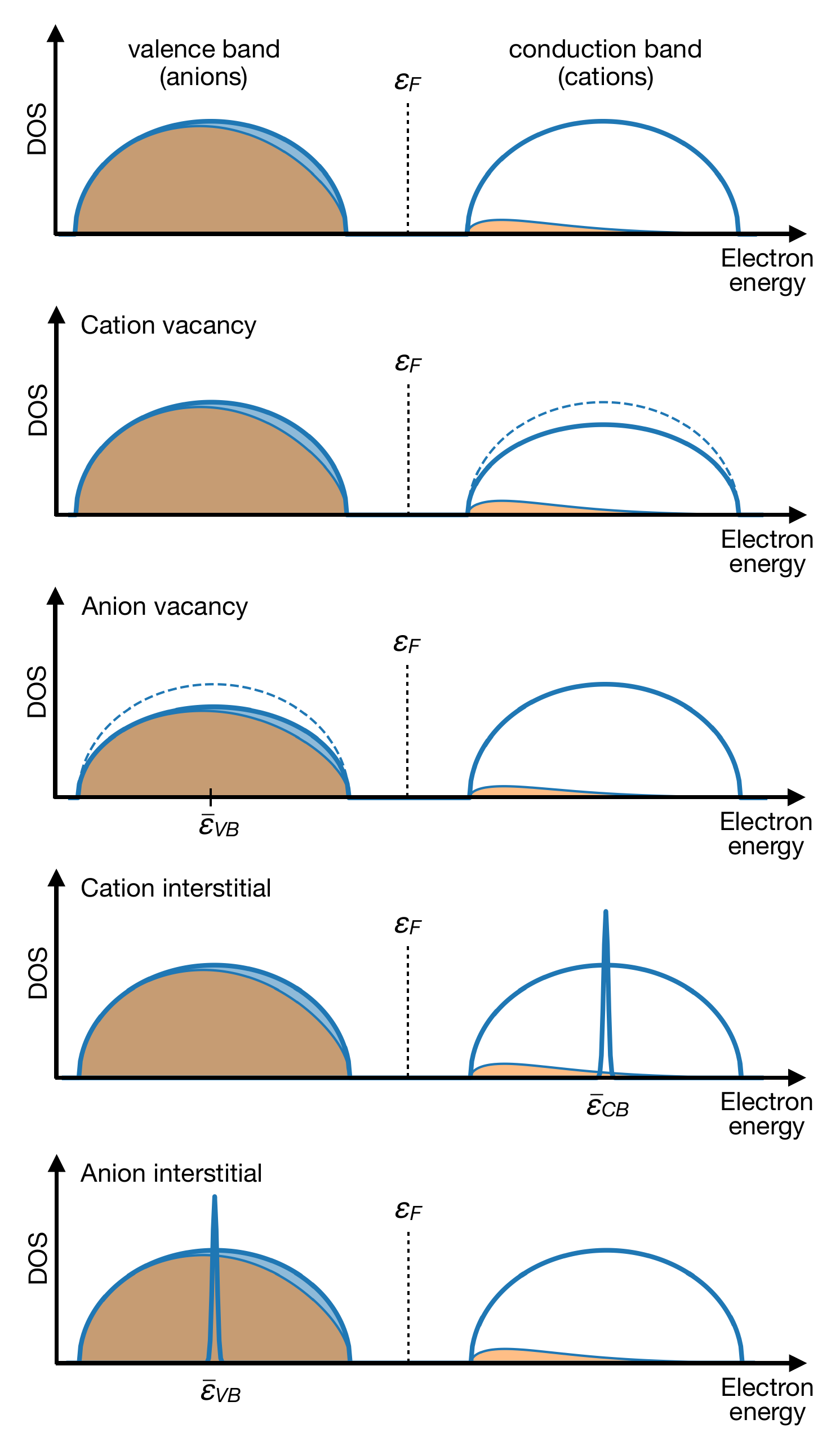}
\caption{\label{fig:3}
Schematics of the electronic density of states (DOS) assumed in our model (topmost panel) and the changes in the electronic DOS due to the formation of various types of vacancy and interstitial defects (lower panels). The shaded regions (in yellow) in the figure represent the occupancy of the bands, based on the Fermi-Dirac statistics.}
\end{figure}

%
%
Let's consider a binary C$_1$A$_1$ ionic semiconductor defined as follows:
\begin{itemize}
\item[({\it i})] it is composed of two kinds of atoms A and C, one more electronegative taking a role of an anion (A) and the other less electronegative (cation-C),
\item[({\it ii})] every anion (cation) contributes $n_a$ ($n_c$) number of atomic orbitals and $N_a$ ($N_c$)  number of electrons where the charge balance between the cations and anions implies $N_a+N_c=n_a$,
\item[({\it iii})] electrons interact only with the nuclei and not among themselves (independent electron approximation),
\item[({\it iv})] band gap forms between two bands, the valence band that is predominantly of the anion character and the cation derived conduction band (Fig.~\ref{fig:3} topmost panel),
\item[({\it v})] defects such as anion or cation vacancies only affect the valence band (anion defects) or conduction band (cation defects) densities of states
, while interstitial defects add atomic orbitals with the energy that falls in the middle of the corresponding band (Fig.~\ref{fig:3} lower panels), and
\item[({\it vi})] we also neglect any changes to the density of states that are due to relaxations of atomic positions upon defect formation.
\end{itemize}

%

For the sake of simplicity, we have intentionally neglected the possibility of vacancy states occurring deep inside the band gap. This is not a big limitation as the deep donors are typically fully ionized close to the top valence band, as are deep acceptors close to the bottom of the conduction band. Under these circumstances the above assumptions should still apply.
To describe this (idealized) binary system, we will utilize a model Hamiltonian with electrons described within the tight-binding approximation and the repulsion between nuclei through the pair potential term \cite{chern_PRL:2017}.
For our discussion it is more useful to write the total energy of the system:
\begin{equation}\label{eq:tot_energy}
E_{tot} = \sum_{k} \varepsilon_k f(\varepsilon_k,T) +  \frac{1}{2} \sum_{i,j} V_{ij},
\end{equation}
where the summation in the first term goes over all occupied electronic states as determined by the Fermi-Dirac distribution function $f(\varepsilon_k,T)$, and the second term is the nuclear repulsion term. Note that this expression is only valid if the electron-electron interactions are neglected.

\subsection{Defect formation energies}\label{ssec:defects}
%
As already discussed, the intrinsic aspects of semiconductor dopability can be formulated in terms of the energy to form intrinsic compensating defects that prevent (compensate) introduction of free charge carriers of the desired type ($n$ or $p$). The energy to form a single point defect is given as:
\begin{equation}\label{eq:dfe}
\Delta E_{D} = E_{tot}^{Defect} - E_{tot}^{Host} \pm \mu,
\end{equation}
where $E_{tot}^{Defect}$ and $E_{tot}^{Host}$ represent the energy of the system with and without the defect $D$, respectively; while $\mu$ is the chemical potential of the respective chemical reservoirs with which the exchange of atoms occurs upon forming the defect. The plus sign in $\pm \mu$ corresponds to vacancies while the minus is used in case of interstitials.
 
{\it Cation vacancy.} Within the above model, formation of a cation vacancy will result in: ({\it i}) absence of a nucleus at a particular cation site $\alpha$, ({\it ii}) removal of the $N_c$ number of electrons from the system, and ({\it iii}) reduction of the density of states primarily in the conduction band in the amount equal to the number of states $n_c$ each cation contributes to the system. As illustrated in Fig.~\ref{fig:3}, we will assume that this reduction in the number of conduction band states is distributed in a way that does not affect the average energy of conduction band while the overall number of states is reduced by $n_c$. It is also important to note that the Fermi statistics in combination with the largely unchanged DOS in the valence band implies that the $N_c$ number of holes created by removing one cation will thermalize over both valence and conduction bands and will have the energy equal to the negative electron chemical potential or the Fermi energy ($\varepsilon_F$) as illustrated in Fig.~\ref{fig:3}. By incorporating these considerations into equations \eqref{eq:tot_energy} and \eqref{eq:dfe} the energy to form one cation vacancy ($\Delta E_{V_c}$) becomes:
\begin{equation}\label{eq:vc}
\Delta E_{V_c} = -N_c \varepsilon_F - V_{\alpha} + \mu_c,
\end{equation}
where the first term on the righthand side represents the energy contribution due to the difference in occupation of the electronic states between the defect and host system ($N_c$ holes with energy $-\varepsilon_F$), the second $V_{\alpha}=\sum_{j} V_{\alpha,j}$ term represents the energy difference in the nuclear repulsion due to the missing cation at the site $\alpha$ (nuclear repulsion potential of that site), and as before, the last term represents the cation chemical potential. In the derivation of eq.~\eqref{eq:vc} we also assumed that creating a cation vacancy does not appreciably affect the Fermi energy, {\it i.e.}, the $\varepsilon_F$ of the defect system equals that of the host. Rigorously, this assumption is valid in the case of low vacancy concentration (the dilute limit). Given the chemical composition and the crystal structure, which set the $V_{\alpha}$ and the some value of the cation chemical potential set by the state of the cation reservoir, the $\Delta E_{V_c}$ becomes the decreasing function of the Fermi energy as in depicted in Fig.~\ref{fig:2}.

As usually done in defect calculations \cite{Lany_2009, freysoldt_RMP:2014}, one can further separate the total chemical potential $\mu_c= \mu_c^o + \Delta \mu_c$ into that of the standard cation phase at standard conditions ($\mu_c^o$), {\it i.e.}, that of the solid metal at room temperature and ambient pressure, and the deviation from the reference value ($\Delta \mu_c<0$). Applying the same model Hamiltonian and the total energy formula from eq.~\eqref{eq:tot_energy} one can write the reference chemical potential as $\mu_c^o=N_c \bar{\varepsilon}_c + \bar{V}_c$, where $\bar{\varepsilon}_c$ represents a characteristic (average) energy of an electron in the reference phase while $V_c$ stands for the average nuclear repulsion energy (per nucleus). While we will keep $\bar{\varepsilon}_c$ as a separate entity, it is worthwhile noting that one can think of $\bar{\varepsilon}_c$ as approximately at the same energy as the center of the conduction band. After implementing these equations for the chemical potential the formation energy of the cation vacancy becomes:
\begin{equation}\label{eq:vc2}
\Delta E_{V_c} = N_c ( \bar{\varepsilon}_c - \varepsilon_F) + (\bar{V}_c - V_{\alpha}) + \Delta\mu_c.
\end{equation}
This equation provides a relatively simple physical picture of the cation vacancy formation, which involves transfer of $N_c$ electrons from the system into the reference phase quantified by the energy of that transfer (first parenthesis) and transfer of a cation nucleus from the system into the reference phase quantified by the difference in the nuclear repulsion (second parentheses). The last term describes the deviation of the actual cation reservoir from the standard state and is a function of the parameters such as temperature and pressure. Lastly, it is important to note that $ \bar{\varepsilon}_c$ and $\varepsilon_F$ need to be expressed relative to the common reference, which is usually assumed to be the vacuum level. Please note that Varley et al.\cite{Varley2017} recently showed that for the tetrahedrally bonded semiconductors the cation vacancy formation energy can be correlated to the branch point energy.

{\it Anion vacancy.} Analogously, in our model the creation of an anion vacancy will result in: ({\it i}) the absence of a nucleus at a 
particular anion site (let's label it $\alpha$ again), ({\it ii}) removal of the $N_a$ number of electrons from the system, and ({\it iii}) reduction of the density of states primarily in the valence band, in the integral amount equal to the number of states $n_a>N_a$ each anion contributes to the system. One can derive an equation for the formation energy of an anion vacancy similar to eq.~\eqref{eq:vc2} with one key difference. Now, the reduction in the number of valence band states needs to be taken into account. The easiest way is to see how to include the reduction in the valence band DOS is to assume low temperature relative to the band gap so that all valence band states are approximately fully occupied. If this is the case then the difference in the summation over all occupied states will amount to the number of states that are missing ($n_a$) times the average energy of the valence band states ($\bar{\varepsilon}_{VB}$). Also, the $N_c$ electrons that were previously occupying the missing states in the valence band will now be thermalized across the band gap giving rise to the $N_c \varepsilon_F$ term. Taking all these contributions into account the anion vacancy formation energy is given as:
\begin{equation}\label{eq:va}
\Delta E_{V_a} = N_c \varepsilon_F - n_a  \bar{\varepsilon}_{VB} -  V_{\alpha} + \mu_a.
\end{equation}
After applying the already described procedure for $\mu_a$, and noting that $N_a + N_c=n_a$ the anion vacancy formation energy becomes:
\begin{eqnarray}\label{eq:va2}
\Delta E_{V_a} &=&N_c (\varepsilon_F -  \bar{\varepsilon}_{VB} ) + N_a ( \bar{\varepsilon}_{a}  -  \bar{\varepsilon}_{VB} ) \nonumber \\
  &+&(\bar{V}_a - V_{\alpha}) + \Delta\mu_a,
\end{eqnarray}
where, as before, $\bar{\varepsilon}_{a}$ represents some characteristic electronic energy of the anion reference phase (close to the center of the valence band), and $\bar{V}_a$ stands for its average nuclear repulsion. Hence, the formation of an anion vacancy requires energy which is an increasing function of $\varepsilon_F$ and involves transfer of the cation electrons ($N_c$) from the valence band to the $\varepsilon_F$, transfer of the anion electrons ($N_a$) from the valence band into the anion reference phase, transfer of the anion nucleus from the system into the anion reference phase, and, as before, the last term describes the deviation of the actual anion reservoir from the standard state.  

{\it Interstitial defects.} 
To first order the interstitial defects, either the cation or the anion ones, can be approximated as contributing their atomic orbitals and their valence electrons to the system. The cation interstitial will contribute the $n_c$ number of localized states at approximately the mid-conduction band energy and the $N_c$ number of electrons that would follow the Fermi-Dirac distribution and distribute themselves within the valence and the conduction bands  so that their average energy in the dilute limit will become $\varepsilon_F$ as illustrated in Fig.~\ref{fig:3}. The anion interstitial on the other hand, contributes $n_a$ partially filled orbitals with $N_a$ electrons at the mid-valence band. Because of the charge transfer to fill these states there will be $N_c=n_a-N_a$ holes created having the energy $\varepsilon_F$. The formation of these two defects requires the energy that can be derived in a similar fashion like previous two amounting to:
\begin{eqnarray}\label{eq:int}
\Delta E_{I_c} &= &N_c(\varepsilon_F -  \bar{\varepsilon}_{c} ) + (V_{\alpha}-\bar{V}_c) - \Delta\mu_c, \nonumber\\
\Delta E_{I_a} &= &N_a(\bar{\varepsilon}_{VB} - \bar{\varepsilon}_{a}) + N_c (\bar{\varepsilon}_{VB} - \varepsilon_{F})\nonumber\\
 &+&(V_{\alpha}-\bar{V}_a) - \Delta\mu_a,
\end{eqnarray}
where $\Delta E_{I_c}$ and $\Delta E_{I_a}$ stand for the formation energy of the cation and anion interstitial defects, respectively, occupying lattice sites that are labeled $\alpha$ in both cases.

%
\subsection{Dopability metrics and the emerging design principles}\label{ssec:dopability}
%
The relevance of the previous derivations to the question of dopability, the very topic of this paper, follows from the Zunger's formulation of dopability \cite{zunger_APL:2003} in terms of the intrinsic compensating defects and the corresponding doping pinning energies $\varepsilon_F^{(n)}$ and $\varepsilon_F^{(p)}$, as already described (see Fig.~\ref{fig:2}). The usual culprits preventing (compensating) the introduction of electrons into the conduction band are intrinsic acceptor defects such as the cation vacancies and/or anion interstitials, while the anion vacancies and cation interstitials typically obstruct $p$-type doping. Having this in mind the expressions for $\varepsilon_F^{(n)}$ and $\varepsilon_F^{(p)}$, in Fig.~\ref{fig:2}, can be defined as:
\begin{eqnarray}\label{eq:pinning}
\varepsilon_F^{(n)} & = & min \{ \varepsilon_F^{(V_c)},\varepsilon_F^{(I_a)},\dots \}, \nonumber \\ 
\varepsilon_F^{(p)} & = & max \{ \varepsilon_F^{(V_a)},\varepsilon_F^{(I_c)},\dots \}.
\end{eqnarray}
where the $n$-type doping limit $\varepsilon_F^{(n)}$ is the minimal doping pinning energy among all the intrinsic acceptor defects, while the $p$-type doping limit $\varepsilon_F^{(p)}$ is the maximal doping pinning energy among all the intrinsic donors. 

To simplify the discussion we will focus on vacancies because in binary systems they are most often the doping limiting defects, and will discuss how the main results changes in case interstitial defects are the limiting factor. Also, to make the expressions easier for discussion, we will express pinning energies relative to the corresponding band edge as $\varepsilon_F^{(n)}=CBM + \delta \varepsilon_F^{(n)} $ and $\varepsilon_F^{(p)}= VBM - \delta \varepsilon_F^{(p)}$ as shown in Fig.~\ref{fig:2}. Under the assumption that vacancies determine the dopability of materials, $\delta \varepsilon_F^{(n)}$ and $\delta \varepsilon_F^{(p)}$ can be derived from the condition that the formation energy of the corresponding defect equals to zero at the pinning energies. From equations \eqref{eq:vc2} and \eqref{eq:va2} one finds:
\begin{eqnarray}\label{eq:metrics}
\delta \varepsilon_F^{(n)} & = & \bar{\varepsilon}_c - CBM + \frac{1}{N_c} (\bar{V}_c - V_{\alpha}) + \frac{1}{N_c} \Delta\mu_c, \nonumber\\
\delta \varepsilon_F^{(p)} & = & VBM -  \bar{\varepsilon}_{VB} + \frac{N_a}{N_c} ( \bar{\varepsilon}_{a}  -  \bar{\varepsilon}_{VB} )  \nonumber \\
 & + &  \frac{1}{N_c} (\bar{V}_a - V_{\alpha}) +  \frac{1}{N_c} \Delta\mu_a.
\end{eqnarray}

\begin{table*}[t]
\caption{\label{tab:1} List and description of various terms appearing in our dopability models is given together with the list of physical quantities and/or proxies employed in the model validation.}
\begin{tabular}{lll}
\hline\hline\\[-0.5em]
Term & Description & Quantity or Proxy used in validation (symbol) \\[0.5em]
\hline\\[-0.5em]
$\bar{\varepsilon}_c$		&average electronic energy in cation reference phase			&work function of cation reference phase ($W_c$)	\\[0.2em]
$\bar{\varepsilon}_a$		&average electronic energy in anion reference phase			&work function of anion reference phase ($W_a$)	\\[0.2em]
$\bar{V}_c - V_{\alpha}$		&difference in nuclear repulsion between cation 				&compound's enthalpy of formation ($\Delta H_{f}$)	\\[0.2em]
						&reference phase and site $\alpha$ in the compound			&										\\[0.2em]
$\bar{V}_a - V_{\alpha}$		&difference in nuclear repulsion between anion 		 		&compound's enthalpy of formation ($\Delta H_{f}$)	\\[0.2em]
						&reference phase and site $\alpha$ in the compound			&										\\[0.2em]
$\bar{\varepsilon}_{VB}$		&average energy of valence band states						&calculated from density of states				\\[0.2em]
$\bar{\varepsilon}_{c}^{s}$	&average energy of cation $s$-states						&calculated from density of states				\\[0.2em]
$CBM$					&conduction band minimum								&GW calculated absolute $CBM$				\\[0.2em]
$VBM$					&valence band maximum									&GW calculated absolute $VBM$				\\[0.2em]
\hline\hline		
\end{tabular}
\end{table*}

Within this model the $\delta \varepsilon_F^{(n)}$ and $\delta \varepsilon_F^{(p)}$ effectively become $n$- and $p$- dopability metrics. The condition for ambipolar dopability requires both pinning energies to be deep inside the corresponding band, which then translates into requiring both $\delta \varepsilon_F^{(n)}$ and $\delta \varepsilon_F^{(p)}$ to be positive and large. The following design principles emerge from these requirements.

{\it Design principles for n-type dopability limited by the cation vacancies.} First, the CBM of the material needs to be as low as possible, however, not relative to some global reference such as the universal pining energy or the branch point energy, but relative to the characteristic electronic energy of the cation reservoir so to maximize the $\bar{\varepsilon}_c - CBM$ term. If one thinks of $\bar{\varepsilon}_c$ as close in energy to the center of the conduction band then maximization of the first term implies large conduction bandwidth. Second, to make $(\bar{V}_c - V_{\alpha})/N_c$ large and positive the nuclear repulsion in the cation reservoir needs to be larger than that in the compound and if this is the case one wants $N_c$ to be small (cation valency). Inversely, if $\bar{V}_c - V_{\alpha}<0$ the $n$-type dopability would require $N_c$ to be large so to minimize the harmful influence of this term. However, to accurately assess the influence of this term one needs to know the actual pair potential for a particular system. We will come back to this question later. 

{\it Design principles for p-type dopability limited by the anion vacancies.} Maximization of the $\delta \varepsilon_F^{(p)}$ implies qualitatively different requirements. Maximizing the first term demands the VBM to be above the mean valence band energy ($\bar{\varepsilon}_{VB}$). This is always true, but this term is large only in systems with large valence bandwidths. Alternatively, one could imagine increasing $VBM -  \bar{\varepsilon}_{VB}$ by having additional valence bands located above the anion band. While not captured by the original model, this will be discussed later in the text. The second term is likely very small because, to first order, one can think that $\bar{\varepsilon}_{a} \approx \bar{\varepsilon}_{VB}$. Much like before, one would like to have $N_a/N_c$ large or small depending on the sign of the energy difference. The third term is analogous to the one for $n$-dopability and demands the nuclear repulsion to be higher in the anion reservoir than in the actual system weighed by the $1/N_c$ term. 

The meaning of the $\Delta \mu$ is the same for both $n$- and $p$-dopability. Since $\Delta \mu \leq 0$, it is desired to have synthesis conditions to be as close to $\Delta \mu = 0$ as possible. In other words, one wants to be as rich in the corresponding element as possible to prevent the formation of its vacancies. In binary systems $\Delta \mu = 0$ is typically a boundary of the stability region both for cations and anions. We will therefore assume in the reminder of this paper that $\Delta \mu = 0$ condition can always be fulfilled. For the purpose of comparing the model with the explicit defect calculations this is a fair assumption, because $\Delta \mu$ term is same in both the cases, and hence, exactly cancels out. However, one needs to be cognizant of the fact that in chemical systems with many competing phases, $\Delta \mu = 0$ might not be possible to achieve for all phases and for both cations and anions, and that the $\Delta \mu$ term may appear as a limiting factor to dopability. But, even in such scenarios, actual range of $\Delta \mu$, based on phase stability analysis, can be accommodated in the model, to provide reasonable estimates for $\delta \varepsilon_F^{(n)}$ and $\delta \varepsilon_F^{(p)}$.

The design principles change in case of interstitials. Similar analysis (see the supplementary for equations) shows that if the donor behaving cation interstitial is the lowest energy defect close to the VBM, then maximizing $p$-type dopability would require: ({\it i}) that the VBM is as high as possible relative to the $\bar{\varepsilon}_c$, and ({\it ii}) that nuclear repulsion at the interstitial site $V_{\alpha}$ is much larger than $\bar{V}_c$ together with relatively small $N_c$ or in the case $V_{\alpha}<\bar{V}_c$ the $N_c$ needs to be large. In case of the anion interstitials the $n$-dopability would benefit from: ({\it i}) small valence band widths and small band gaps, ({\it ii}) anions with low-lying atomic orbitals, and ({\it iii}) high nuclear repulsion on the interstitial site.

As evident from this discussion, the dopability of semiconductors does not depend on a single material property and is a product of relatively complex tradeoffs between different properties. Also, as previously noted the interstitial defects are usually higher in energy than vacancies in the simple binary systems considered here. This is particularly true for the anion interstitials due to their ionic sizes requiring large amount of space. This is why in the reminder of the text we will focus on vacancies as dominant dopability limiting factors. Only one compound in our study, Mg$_2$Si, turns out to have its $p$-type dopability limited by the cation (Mg) interstitials, for which we do consider the described interstitial defect model.
%
\section{Validation}\label{sec:validation}
%
Validation of our model is done by comparing how accurately it reproduces numerical values for the dopability metrics, $\delta \varepsilon_F^{(n)}$ and $\delta \varepsilon_F^{(p)}$, which we compute directly using modern defect theory and defect calculations. However, the quantities appearing in eq.~\eqref{eq:metrics} are not easily accessible causing direct evaluation of the model dopability metrics difficult. This is particularly the case for the average electronic energies of the elemental reservoirs $( \bar{\varepsilon}_c,  \bar{\varepsilon}_a)$ as well as the ionic repulsion terms $(V)$ for both the elemental reservoirs and the material of interest. The approach we adopt here is to find proxies for these hard-to-access quantities instead of trying to calculate them directly. This approach follows the same idea as in our previous work in which we successfully developed models for electronic and heat transport in thermoelectric materials using more accessible proxies instead of hard-to-compute quantities \cite{gorai_NRM:2017}. If appropriate physical proxies can be found, a simplified model involving these proxies can be made computationally inexpensive as well as predictive. The price of such an approach, however, is the introduction of free parameters into the model that need to be fit to the existing data. The performance of the model is then assessed by the quality of the fit. The following discussion focuses on the dopability metrics derived assuming vacancies as the lowest energy compensating defects. Dopability metrics based on interstitials defects are also derived in the supplementary information.

{\it $n$-type dopability metric.} As shown in eq.~\eqref{eq:metrics} there are three main contributions to $\delta \varepsilon_F^{(n)}$, the electronic term, nuclear repulsive term, and the chemical potential term. The electronic term $\bar{\varepsilon}_c - CBM$ involves average electronic energy of the cation reference phase (solid metal) and the energy of the conduction band minimum. An intuitive proxy for $\bar{\varepsilon}_c$, that is easily accessible from literature, is the negative work function ($-W_c$) of the reference phase. However, the work function of metals if expressed as an energy relative to vacuum (negative value) represents the maximal electronic energy and not an average one. Therefore, to account for the difference between $\bar{\varepsilon}_c$ and work function, we consider replacing the first term with the proxy of the form $a \times (-W_c) - b \times CBM$, where $a$ and $b$ are the fitting constants, $W_c$ is the work function of the metal phase and $CBM$ the conduction band minimum (negative electron affinity) of the semiconductor material (measured or calculated). Both $W_c$, and $CBM$ are expressed relative to the vacuum. 

%
\begin{figure}[!t]
\includegraphics[width=\linewidth]{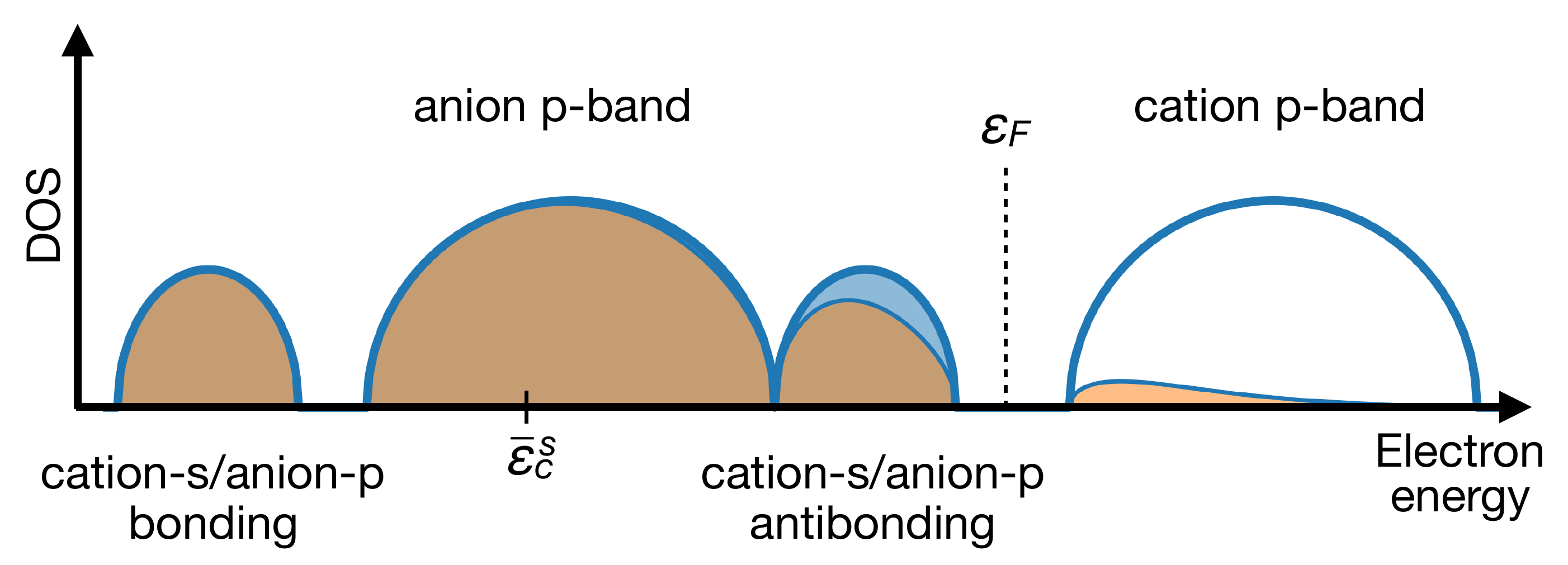}
\caption{\label{fig:4} Schematic of the electronic DOS of binary systems with low valent cations such as the group-IV and group-V chalcogendes (SnO, SnS, PbSe, PbTe, Bi$_2$Se$_3$, Bi$_2$Te$_3$). The ``center of mass'' of the the filled cation-$s$ contributions to the DOS is denoted as $\bar{\varepsilon}_{c}^{s}$.}
\end{figure}

To account for the nuclear repulsion contribution one could fit the parameters of our tight-binding plus nuclear repulsion Hamiltonian to a set of materials properties such as the equilibrium volume, cohesive energy, bulk modulus, etc. 
While this would be the most appropriate thing to do, it would need to be done for all materials of interest and all elemental phases which would render the whole process impractical. Among material properties, cohesive energies, volume and bulk modulus are well known to be correlated with each other \cite{Verma2006,Wacke2011}. Hence simply considering cohesive energy as a proxy for the nuclear repulsion contribution for a material could suffice. However, as we are interested in the difference ($\bar{V}_c - V_{\alpha}$) between the compounds and the elemental phase energies, we chose $\Delta H_f$ as a proxy as it already include the difference between the compounds and the elemental phase energies, and replace the nuclear repulsion term with linear dependence on the compounds enthalpy of formation $c \times \Delta H_f + d$. Admittedly,  $\Delta H_f$ includes contributions both from electrons and nuclei. The choice is motivated by the physical relation to the nuclear repulsion and, in part, by our previous work in which we discovered the relevance of $\Delta H_f$ to the formation energy of oxygen vacancies in metal oxides \cite{deml_JPCL:2015}. The $n$-type dopability is evaluated under cation-rich condition such that $\Delta \mu_c=0$, because it represents the most favorable thermodynamic condition to dope the semiconductor $n$-type.

Lastly, to extend the model to systems with partially oxidized cations such as the group-IV and group-V chalcogenides such as SnO, SnS, PbSe, PbTe, Bi$_2$Se$_3$, Bi$_2$Te$_3$, one needs to include an additional term to account for the filled cation $s$-states that contribute to the valence band as shown in Fig.~\ref{fig:4}. This is easily done by including the  transfer of $N_c^s$ number of the cation $s$-electrons from the valence band to the cation reservoir upon forming the cation vacancy. This term can be written as $N_c^s (\bar{\varepsilon}_c - \bar{\varepsilon}_c^s$) where, $\bar{\varepsilon}_c$ is the average electronic energy of the cation reference phase while $\bar{\varepsilon}_c^s$ is the average energy of the cation $s$-states in the material's valence band (s-DOS center of mass). The latter is obtained from the bulk electronic structure calculations, similar to the average energy of the anion $p$-states in the $p$-type dopability metric. The $n$ dopability metric, formulated in terms of the proxies is now given as:
\begin{eqnarray}\label{eq:n_proxy}
\delta \varepsilon_F^{(n)} & = & a^{(n)} \times W_{c}  - b^{(n)} \times CBM - c^{(n)} \times \bar{\varepsilon}_{c}^{s} \nonumber\\
& + & d^{(n)} \times \Delta H_f + e^{(n)},
\end{eqnarray}
where $a^{(n)}, b^{(n)}, c^{(n)}, d^{(n)}, e^{(n)}$ are the free parameters of the model (fitting constants), that are obtained by fitting to the directly calculated $\delta \varepsilon_F^{(n)}$ from first-principles defect calculations. The above form also allows extending the model to any C$_x$A$_y$ stoichiometry (not only C$_1$A$_1$). The list of terms appearing in the models as well as the corresponding proxies we use for validation is given in Table~\ref{tab:1}

{\it $p$-type dopability metric.} Analogously, terms such as VBM, $\bar{\varepsilon}_a$, and ($\bar{V}_a - V_{\alpha}$) in the definition of $\delta \varepsilon_F^{(p)}$ are substituted with the valence band maximum (negative ionization potential) of the semiconductor, negative work function of the anion reference phase or in case of molecules negative first ionization potential ($-W_a$), and the compound's enthalpy of formation. As discussed earlier, $p$-type dopability metric has an additional term ($\bar{\varepsilon}_{VB}$) representing average energy of the valence band, which we evaluate for a set of classic binary semiconductors as the average energy of the anion-$p$ states. This is done from the electronic structure calculations (see the methods section) of the bulk, defect-free materials. Finally, the $p$-type dopability is evaluated under anion-rich conditions such that $\Delta \mu_a=0$. The $p$ dopability metric, formulated in terms of proxies is given as:
\begin{eqnarray}\label{eq:p_proxy}
\delta \varepsilon_F^{(p)} & = & a^{(p)} \times W_{a}  + b^{(p)} \times VBM - c^{(p)} \times \bar{\varepsilon}_{VB} \nonumber\\
& + & d^{(p)} \times \Delta H_f + e^{(p)},
\end{eqnarray}
where, as before, $a^{(p)}, b^{(p)}, c^{(p)}, d^{(p)}, e^{(p)}$ are free parameters that are fitted to the directly calculated $\delta \varepsilon_F^{(p)}$. For each $\delta \varepsilon_F^{(n)}$ and $\delta \varepsilon_F^{(p)}$ we have a total of 5 fitting constances which we fit to a set of 16 materials as described further.

The work functions for the cation reference phases, the ionization energies for the gaseous species such as O$_2$, N$_2$ and the enthalpy of formations are obtained from the Refs.\cite{crchandbook, NIST_ASD}, with details provided in Table~S5 of the supplementary information (SI). Compounds' conduction band minima and valence band maxima (with respect to vacuum) are explicitly calculated using the standard methodology, involving the combination of GW electronic structure and DFT surface calculations \cite{vladan_PCCP:2014}. Experimental $CBM$ and $VBM$ values are used for Ga$_2$O$_3$ and In$_2$O$_3$. Details of all the above intrinsic material properties of the compounds, along with the average energy of the anion-$p$ and cation-$s$ states (obtained from calculated DOS) employed in our model are provided in Table~S6 of the SI.

\begin{figure}[!t]
\includegraphics[width=\linewidth]{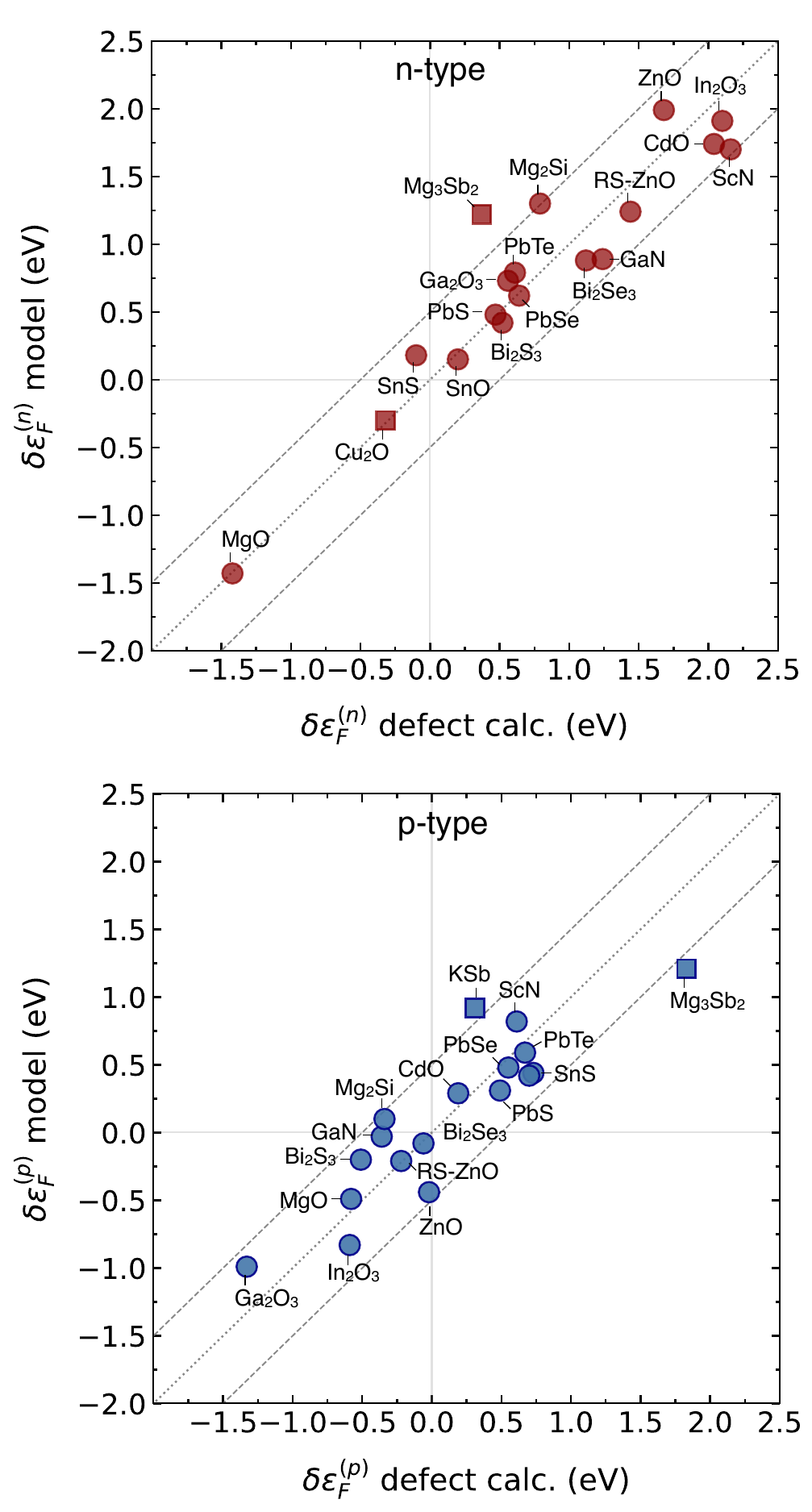}
\caption{\label{fig:5} Comparison of the analytic model and the calculated (from first-principles defect calculations) $n$- and $p$-type dopability metrics $\delta \varepsilon_F^{(n)}$ and $\delta \varepsilon_F^{(p)}$. The model parameters from eqs.~\eqref{eq:n_proxy} and \eqref{eq:p_proxy} are obtained via linear regression to the calculated values. Mg$_3$Sb$_2$, KSb and Cu$_2$O (shown as squares) are employed for model validation and are not included in the fitting. See text for more details.}
\end{figure}

We performed defect calculations for a set of 16 binary ionic-semiconductors including classic III-Vs and II-VIs, group-III oxides, and lead and bismuth chalcogenides. Results of the fitting of the $n$ and $p$-type dopability metrics eqs.~\eqref{eq:n_proxy} and \eqref{eq:p_proxy} to the same quantities from defect calculations are shown in Fig~\ref{fig:5}.
We use standard linear regression to obtain values for the fitting parameters. The quality of the fit (as shown in Fig.~\ref{fig:5}) is very good with the R$^2$=0.91 and 0.81, and root mean square error RMSE=0.26 eV and 0.25 eV for $n$ and $p$-type dopability metric, respectively. The final  expressions for the two dopability metrics are:
\begin{eqnarray}\label{eq:fit}
\delta \varepsilon_F^{(n)} & = & 1.40 \, W_{c}  - 0.60 \, CBM + 0.05 \,\bar{\varepsilon}_{c}^{s} + 0.03 \, \Delta H_f \nonumber\\
& + & 4.79, \nonumber\\
\delta \varepsilon_F^{(p)} & = & 0.09 \, W_{a}  - 0.03 \, VBM + 0.37 \, \bar{\varepsilon}_{VB} - 0.12 \, \Delta H_f \nonumber\\
& + & 2.81.\nonumber\\
\end{eqnarray}
%
%
%
%
\begin{figure*}[!t]
\includegraphics[width=0.98\linewidth]{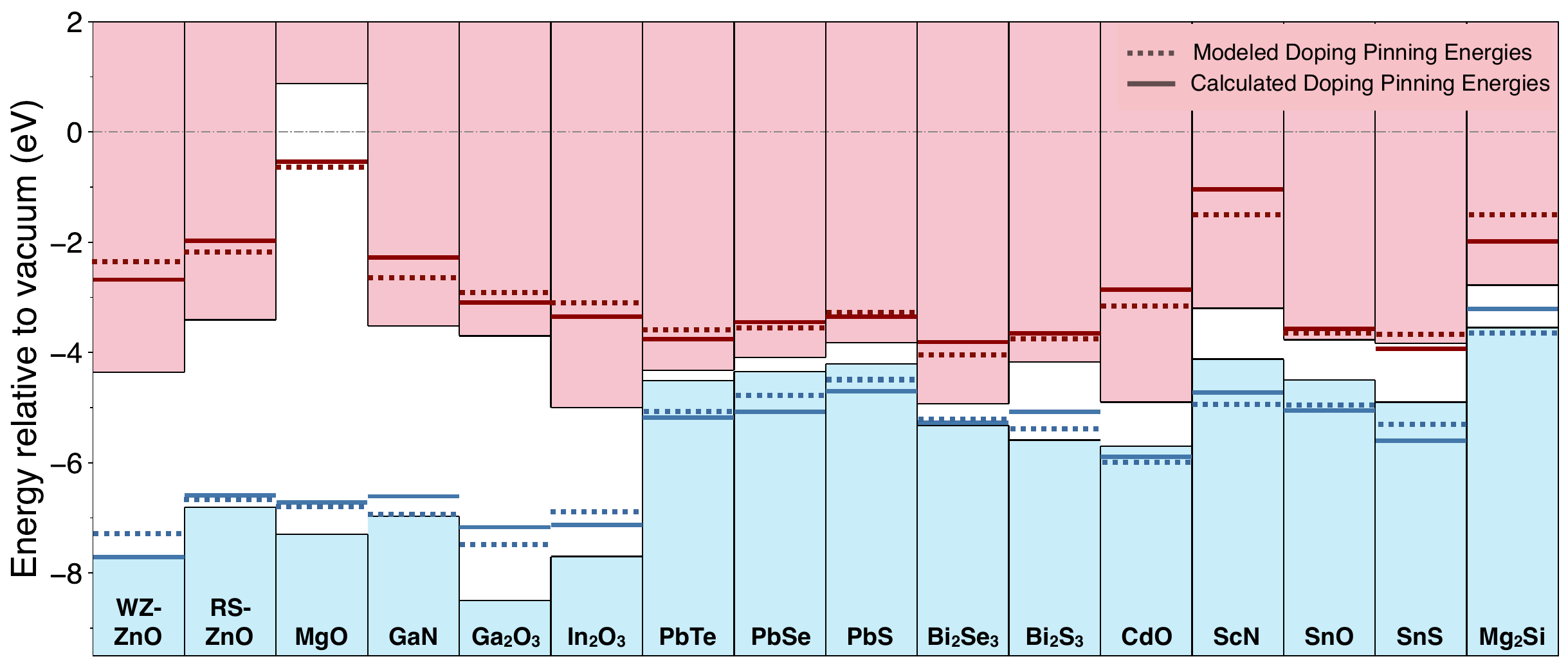}
\caption{\label{fig:6} The doping pinning energies both from direct defect calculations (solid lines) and from our models (dashed lines) are shown together with the absolute band edge position (colored rectangles) for all 16 materials considered in this work.}
\end{figure*}

In addition to relying only on the quality of the fit, our model from eq.~\eqref{eq:fit} was further validated against three additional compounds, Mg$_3$Sb$_2$, KSb, Cu$_2$O that were not included in the fitting. As shown in Fig.~\ref{fig:5}, both $p$- and $n$-type dopability fits very well for Mg$_3$Sb$_2$ because the bonding in Mg$_3$Sb$_2$ satisfies the model assumptions. The $p$-type dopability in KSb fairs well because the valence band is composed of anion (Sb) $p$-states. However, $n$-type dopability for KSb does not agree with the model (not shown), which is not surprising, because the conduction band of KSb does not follow the assumptions of the model, i.e., it is composed of the Sb-$p$ orbitals. The original model also does not capture many transition metal compounds which often have contributions of the metal $d$-states to both valence and conduction bands. However, the for systems such as Cu$_2$O where the conduction band is largely composed of Cu-$s$ states, our model still works well for the $n$-type dopability as illustrated in Fig.~\ref{fig:5}. The $p$-type dopability is not well captured because Cu $d$-states contribute heavily to the valence band but not the conduction band\cite{Lany2015}. That said, the model itself can be modified to include these various cases.

It is important to note that all the modeled values of $\delta \varepsilon_F^{(n)}$ and $\delta \varepsilon_F^{(p)}$ are within 0.5 eV compared to the values from the defect calculations, which is in our opinion striking given the simplifications and approximations adopted in the model. The magnitude of the coefficients and the contribution of each term in eq.~\eqref{eq:fit} vary with the dopability type. For the $n$-type pinning energy, $\delta \varepsilon_F^{(n)}$, individual terms comprising of $W_c$, $CBM$ and the intercept term, contribute more significantly because they are about an order of magnitude larger than the terms comprising of $\bar{\varepsilon}_{c}^{s}$ and $\Delta H_f$. However, for the $p$-type pinning energy, $\delta \varepsilon_F^{(p)}$, individual terms comprising $\bar{\varepsilon}_{VB}$ and the intercept have an order of magnitude larger coefficients than the $W_a$, $VBM$ and $\Delta H_f$ terms. We will come back to the question of dominant terms in the discussion section. It is also important to note that some of the coefficients turn out to be negative. This is possible because: ({\it i}) when substituting physical quantities for proxies we do not know the actual dependencies between the two, ({\it ii}) by using the work functions we are using maximal electronic energy of the elemental reservoirs rather than the average one, and lastly, ({\it iii}) we are folding into the fitting coefficients both the dependence on the actual stoichiometry and number of valence electrons of cations and anions which could also alter the signs of the coefficients. By doing so we are trying to develop a model that is simple to use. Obviously, the choices we made are not unique and one could in principle come up with the different set of proxies and a different numerical model.

\section{Discussion}\label{sec:discussion}
%
The dopability metrics $\delta \varepsilon_F^{(n)}$ and $\delta \varepsilon_F^{(p)}$ and the resulting doping pinning energies are shown in Fig.~\ref{fig:6} alongside the absolute band edge positions for all 16 materials considered in this study. All energies are shown relative to the vacuum level. The $\varepsilon_F^{(n)}$ and $\varepsilon_F^{(p)}$ from defect calculations are represented as continuous lines while those resulting from the model, eq.~\eqref{eq:fit}, are depicted as dashed lines. The position of the pinning energies relative to the band edges determine the dopability of a material. For example, if the pinning energies $\varepsilon_F^{(n)}$ and $\varepsilon_F^{(p)}$ are both inside the corresponding bands (meaning $\delta \varepsilon_F^{(n)} > 0$ and $\delta \varepsilon_F^{(p)} > 0$), the compound allows both $p$- and $n$-type doping, likely to the degenerate levels. Conversely, if one of the doping pinning energies is inside the band gap, the dopability of the material to the corresponding carrier type is reduced and even diminished depending on the distance from the band edge. If both doping pinning energies are deep inside the band gap, the material cannot be doped. 

As evident from the Fig.~\ref{fig:6} as well as from the previous discussion there is a good overall correspondence between the calculated and modeled pinning energies. However, contrary to the Zunger's finding for III-V and II-VI semiconductors, the doping pinning energies generally do not align. It is clear that if compounds outside these two groups are considered the deviations in the positions of the doping pinning levels are significant ($>$ 2 eV). Hence, one can not rely on the alignment of the pinning levels and a simple doping principles based on the band edges alone. 

Lot of emphasis has been given previously to the absolute position of the band edges as a guiding principle. In addition to the work of Zunger and co-workers \cite{zunger_APL:2003, Zhang2000}, Walukiewicz \cite{Walukiewicz1988} and Schleife et al. \cite{schleife_APL:2009} discussed how the band edge energies expressed relative to a universal branch-point energy correlate with dopability. In all of these works it was found that the $n$-type dopable materials typically have high electron affinity (low $CBM$) while the $p$-dopable ones have small ionization potentials (high $VBM$) \cite{schleife_APL:2009, Robertson2011}.

Absolute $VBM$ and $CBM$ also appear in our model as properties influencing materials dopability. Our derivation explains why they correlate with dopability, but also reveals that these are not the only relevant properties as one can clearly see that the dopability trends do not exactly correspond to the trends in the band edge positions. For example, GaN, Ga$_2$O$_3$ and In$_2$O$_3$ are all degenerately $n$-dopable despite large differences in their $CBM$ positions. Also, SnO and SnS are only moderately $n$-dopable while having their $CBM$s below GaN, etc. An obvious question that follows from this discussion is whether a model of dopability that includes only the band edges can be constructed? 
%
\subsection{Description from the band edges alone} 
%
To answer this, we repeated the fitting exercise from validation section by only including the absolute band edge terms and the free parameter (intercept). The simplified dopability metrics are then given as $\delta \varepsilon_F^{(n)} = -b^{(n)} \times CBM + e^{(n)}$ and $\delta \varepsilon_F^{(p)} = b^{(p)} \times VBM + e^{(p)}$. The fit thus performed resulted in the root mean square errors (RMSE) of 0.67 eV and 0.41 eV for the $n$- and $p$-type dopability, respectively; compared to the RMSE of 0.26 eV ($n$-type) and 0.25 eV ($p$-type) for the full model. So, using only the band edge positions as descriptors of dopability is much less accurate than the full model. 

Including the work functions to the two metrics, $\delta \varepsilon_F^{(n)} = a^{(n)} \times W_{c} - b^{(n)} \times CBM + e^{(n)}$ and $\delta \varepsilon_F^{(p)} = a^{(p)} \times W_{a} + b^{(p)} \times VBM + e^{(p)}$, helps improve the model significantly to RMSE =  0.34 ($n$-type) and 0.31 ($p$-type). This further supports one of the key implications of our model. It is not the absolute $VBM$ and $CBM$ that matter, but the $VBM$ and $CBM$ relative to the average electronic energies of the elements, which are in this case represented by their work functions. That said, the absolute $VBM$ and $CBM$ could serve as coarse guidelines but one needs to be aware that the error in the doping pinning energies can likely be much above 0.5 eV. 

\begin{figure}[!t]
\includegraphics[width=0.95\linewidth]{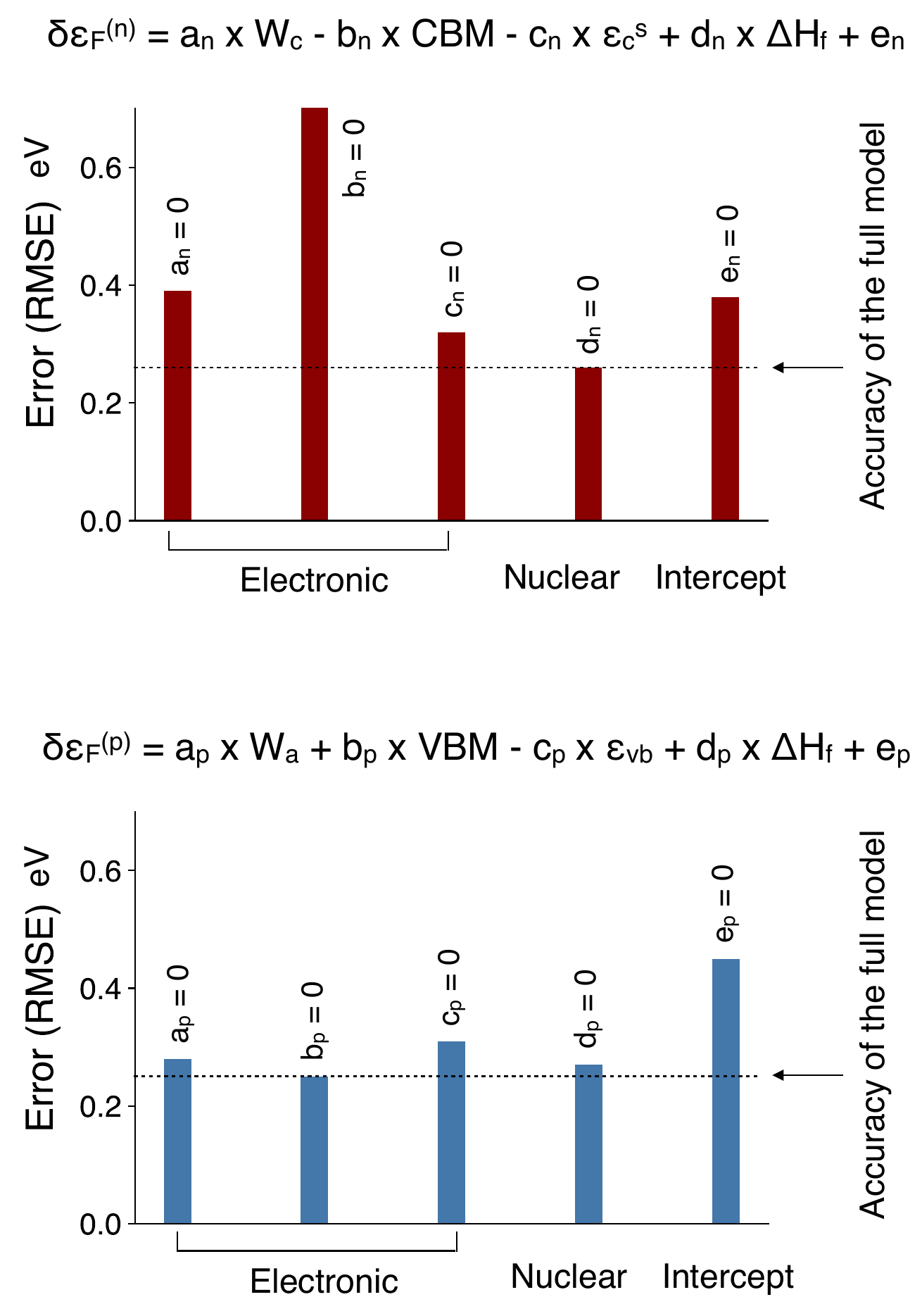}
\caption{\label{fig:7} Comparing the error in the modeled pinning energies when individual terms in Eqs.~\ref{eq:n_proxy} and \ref{eq:p_proxy} are removed one at a time (by setting the respective coefficients to zero)  vs. the full model.}
\end{figure}

Furthermore, to gauge the importance of each individual term, we performed an exercise of removing terms one at a time (by putting their coefficient to zero), and re-doing the fit. As discussed previously, these terms can be grouped into electronic contributions, comprising of reference phase work function ($a \times W$),  absolute band edge positions ($b \times CBM$ or $VBM$), average electronic energy of cation-$s$ ($c \times \bar{\varepsilon}_{c}^{s}$) or valence band states ($c \times \bar{\varepsilon}_{VB}$) and nuclear repulsion ($d \times \Delta H_{f}$),  plus a free parameter (intercept term $e$). Based on the results (RMSE values) summarized in Fig.~\ref{fig:7}, we can draw following conclusions: (1) $n$-type doping metric $\delta \varepsilon_{F}^{(n)}$ depends more significantly on the electronic contributions than the p-type doping metric $\delta \varepsilon_{F}^{(p)}$, (2) within the electronic terms, both the reference phase work function and compounds' CBM are needed to accurately reproduce both $\delta \varepsilon_{F}^{(n)}$ and $\delta \varepsilon_{F}^{(p)}$, and finally, (3)  the intercept is important, especially in case of $\delta \varepsilon_{F}^{(p)}$. Please note, that the intercept integrates both electronic and nuclear contributions that are not easily separated. 

%
\subsection{Role of the band gap} 
Magnitude of the electronic band gap is another quantity used to gauge dopability of semiconductors. It follows from the observation that wider the gap is, more insulating or less dopable a material is likely to be. While generally true, this coarse rule of thumb does not provide any insight into the apparent doping asymmetry (typically towards $n$-type) of many semiconductors. Also, there are exceptions from this rule like MgO for example. MgO is a large band gap ($\sim$7.7 eV) material and is moderately $p$-type dopable \cite{Tardio2002}, which is in agreement with the predicted $p$-doping pinning energy that appear near the $VBM$ in Fig.~\ref{fig:6}. Our approach shows that the band gap of the material does not appear to be a governing factor in determining dopability. What matters are the positions of individual band edges, characteristic electronic energy of the corresponding reference phase, average band energies, band widths, and nuclear repulsions due to creation of a defect. 

The only place where the band gap explicitly appears is in the $n$-type dopability when limited by the anion interstitials (see Dopability metric section. One of the conditions demands that the difference between the center of the valence band and the conduction band minimum $\bar{\varepsilon}_{VB} - CBM$ be as large as possible. Since $\bar{\varepsilon}_{VB} <  CBM$, the only way to accomplish this is to have $CBM$ to be as close to $\bar{\varepsilon}_{VB}$, which implies having narrow valence band width and small band gap, both at the same time. However, since the anion interstitials are rarely low energy defects in binary systems the explicit role of the band gap is not very prominent. 

However, the likelihood for the conditions for the vacancy limited $p$- and $n$-type dopability to be simultaneously fulfilled is higher in low gap systems although not exclusively. Recall that the approximate way to treat average electronic energies of the reference phases is to assume they are close to the conduction and valence band centers. Then, the ambipolar dopability demands large bandwidths for both valence and conduction bands. This is more often the case in narrow gap systems than in wide gap systems, although one cannot neglect exceptions to this relatively naive expectation. There is nothing that in principle forbids moderate to wide gap systems to have large bandwidths and there are also tradeoffs with other properties that could make up for the bands that are not as wide. 
%
\subsection{Extrinsic defects, covalent and multinary systems and the utility in materials searches} 
%
Dopability of a semiconductor can also depend on the availability of an appropriate external dopant. The dopant's effectiveness, in addition to the host material not developing intrinsic compensating defects (the subject of this paper), will depend on: (1) its solubility in the host material, and (2) its preference toward the expected behavior (donation or acceptance of charges). For example, as Fig.~\ref{fig:6} suggests ZnO in its ground state wurtzite structure should be moderately $p$-type dopable. Not to degenerate levels, but nevertheless it should be possible to dope it $p$-type. Based on the solubility among the external dopants, Li (group-I) and N (group-V) are best suited for this purpose. However, Li fails to dope ZnO $p$-type because it is present in nearly equal amount both as a substitutional acceptor (Li$_{Zn}^{1-}$) and as interstitial donor (Li$_{i}^{1+}$). Hence, it almost self-compensates resulting in a very low net hole concentrations. N on the other hand act as a deep acceptor in ZnO, and need high ionization energy to provide any measurable free carriers at room temperature. In ZnO, $p$-type doping is also made difficult because of the presence of hydrogen (acting as a donor) as an unintentional contaminant during growth techniques \cite{Anderson2009, Ozgur2005a, McCluskey2015}.

In principle our model can be extended to include extrinsic defects. This would require knowledge of the position of the atomic orbitals of extrinsic dopants relative to the band edges of the host as well as assumptions (or a model) pertaining to the defects states within the host. Also, the present model is developed for ionic and partially ionic semiconductors. Therefore, it is not expected to apply to fully covalent systems such as the elemental Si, Ge, diamond, and other. In these cases, defects such as vacancies introduce states deep inside the band gap, in addition to renormalizing the valence and conduction band density of states. Such deep defect states can both accept and donate charges, and hence, can limit both the $p$- and $n$-type doping. While the model can be extended to cover fully covalent materials, in this paper we have focused on ionic or partially ionic semiconductors as they constitute a larger group of materials. 

The model description would also become much more complex if one moves beyond the elemental and binary semiconductors. In ternary and multinary systems, generally more than one atomic specie contributes to the band edges, and therefore, it is likely that more than one model is required to cover different possible situations. Alternatively, for specific families among multinary compounds, such as Zintl phases for example, one could try finding some other higher-level descriptors that could be helpful in large scale materials searches \cite{gorai_jmca:2019}.

Lastly, in regard to the large scale materials screening for dopability, the model we have developed could be useful, especially the version involving proxies from the validation section, provided that a rapid evaluation of the absolute (or relative) $VBM$, $CBM$ and elemental work functions can be made possible. While the band edges are not the only governing properties their contributions to the dopability of semiconductors cannot be ignored, not even approximately. The computational procedure to evaluate absolute $VBM$ and $CBM$ is nowadays well established; it includes the electronic structure calculations for the bulk in combination with the calculations of the surface dipoles. The surface calculations represent a real bottleneck to high-throughput screening as they are of the similar level of complexity as the direct defect calculations. So, until a more efficient way to evaluate absolute $VBM$ and $CBM$ position is developed, relatively tedious and laborious direct defect calculations remain the only available choice for a robust predictions of the doping tendencies of materials.   
%
\section{Conclusions}\label{sec:conclusions}
%
In conclusion, in order to reveal the intrinsic materials properties that determine dopability of semiconductors, we have developed a model description of the defect formation energies in ionic and partially ionic systems. The model is constructed using the tight-binding description of the electronic structure augmented by the nuclear repulsion (pair) potential term. Utilizing such an approach in combination with the existing formulation of the dopability in terms of the limiting (compensating) intrinsic point defects, we are able to analytically separate various contributions. Contrary to the presently adopted and largely heuristic view, the position of the band edges alone cannot be used to accurately describe the doping limits of semiconductors. In addition, the electronic structure of the elemental reservoirs has to be taken into account as well as the differences in the nuclear repulsion between the material of interest and the elemental reservoirs. Hence, the dopability of semiconductors is a result of a relatively complex tradeoffs between various intrinsic properties. To make the model practical, as well as for the purpose of validation, we replace the hard-to-calculate quantities in the model with the more accessible ones and showed that model is able to accurately reproduce the directly calculated (from modern defect calculations) doping limits of 16 classic binary semiconductors. Lastly, we discuss the extension of our work to more complex chemistries and the utility in large-scale material searches.

\section{Methods}\label{sec:methods}
{\it First-principles defect calculations.} In this work we employ the standard supercell approach \cite{Lany2009} using our computational framework \cite{Goyal2017} to calculate formation energies of point defects using the following equation:
\begin{equation}
\begin{split}\label{eq:dfe_2}
\Delta H_{\mathrm{D}, q} (E_{F}, \mu) =  & \,\,[E_{\mathrm{D},q} - E_{\mathrm{H}}] + \sum_{i}n_{i} \mu_{i} + \\
&+ qE_{F} + E_{\mathrm{corr}},
\end{split}
\end{equation}
where $\Delta H_{\mathrm{D}, q}$ represents the formation energy of a point defect D in charge state $q$. $E_{\mathrm{D},q}$ and $E_{\mathrm{H}}$ are the total energies of the supercells with and without the defect, respectively. $\mu_{i}$ is the chemical potential of atomic species, $i$, describing exchange of particles with the respective reservoirs. $E_{F}$ is the Fermi energy and is used here to account for the possible exchange of charge between the defect and the Fermi energy (i.e. the charge reservoir). $E_{\mathrm{corr}}$ is a correction term to accounts for the finite-size corrections within the supercell approach \cite{Lany2009}. The chemical potential $\mu_i = \mu_i^{0} + \Delta \mu_{i}$ is expressed relative to the reference elemental chemical potential $\mu_i^{0} $, calculated using the FERE approach \cite{Stevanovic2012} (re-fitted for HSE calculations,), and $\Delta \mu_{i}$  is the deviation from the reference elemental phase, the bounds of which are determined by the thermodynamic phase stability. 

A plane wave energy cutoff of 340 eV, and a Monkhorst-Pack k-point sampling\cite{Monkhorst1976} is used. The low-frequency total (electronic + ionic) dielectric constant is obtained following the procedure in Ref.\citenum{Peng2013}. We have implemented tools in our framework\cite{Goyal2017} to calculate the following finite-size corrections: (1) potential alignment, (2) image-charge correction, and (3) band filling correction to address Moss-Burstein-type effects. Beyond the finite-size effects, another source of inaccuracy arises from the well-known DFT band gap problem. Accurate band gaps are needed to correctly describe the formation energy of charged defects as a function of the electronic chemical potential i.e., Fermi energy. We employ state-of-the-art GW quasiparticle energy calculations\cite{Lars1965} to compute band edge shifts (relative to the DFT-computed band edges). The band edge shifts are used to correct the defect formation energy in multiple charge states. GW calculations are performed on the DFT relaxed structures, with the unit cell vectors re-scaled to match the experimental lattice volume\cite{Peng2013}. For hybrid functional (HSE06\cite{Heyd2003, Heyd2006}) calculations, the the exchange mixing is used accordingly to match their experimental lattice parameters and band gaps. Having defect formation energy allows thermodynamic modeling of defect and carrier concentrations, computed here using the approach from Refs. \cite{Goyal2017b, goyal_PRM:2018, Biswas2009} Confidence in our predictions stems from the correct description of defects and doping in our previous works \cite{Goyal2017b, goyal_PRM:2018, Goyal2017}demonstrating good agreement between calculated and measured defect and charge carrier concentrations in PbTe and other systems.

\begin{suppinfo}
We provide detailed tabulated data of defect formation energies and converged calculation parameters as *.csv files along with the supporting information containing additional details that may be necessary to validate our results and discussion presented in the main manuscript. This material is available at http://pubs.acs.org/.
\end{suppinfo}

\begin{acknowledgement}
The authors thank Dr. Stephan Lany from National Renewable Energy Laboratory (NREL) for fruitful discussions. We acknowledge support from NSF DMR program, grant no. 1729594 and 1729487. This research used computational resources sponsored by the DOE Office of Energy Efficiency and Renewable Energy and located at National Renewable Energy Laboratory. High-performance computational resources at Colorado School of Mines are also acknowledged.
\end{acknowledgement}


\bibliography{biblio}

\end{document}